\documentclass{emulateapj}
\usepackage{natbib}

\shorttitle{9 New M Star Companions}
\shortauthors{Daemgen et al.}

\begin{document}

\title{Discovery of 9 New Companions to Nearby Young M Stars with the Altair AO System}

\author{Sebastian Daemgen\altaffilmark{1}, Nick Siegler\altaffilmark{1}, I. Neill Reid\altaffilmark{2}, \& Laird M. Close\altaffilmark{1}}

\altaffiltext{1}{Department of Astronomy and Steward Observatory, University of Arizona, Tucson, AZ 85721}
\altaffiltext{2}{Space Telescope Science Institute, Department of Physics and Astronomy, Baltimore, MD 21218}
\slugcomment{To appear in the January 1, 2007 issue of the Astrophysical Journal}
\email{sdaemgen@as.arizona.edu}

\begin{abstract}
We present results of a high-resolution, near-infrared survey of 41 nearby, young ($\lesssim\!300\mathrm{\,Myr}$) M0--M5.0 dwarfs using the Altair natural guide star adaptive optics system at the Gemini North telescope. Twelve of the objects appear to be binaries, 7 of which are reported here for the first time. One triple system was discovered. Statistical properties are studied and compared with earlier (F to K) and later ($\ge$ M6 very low-mass, VLM) populations. We find that the separation distribution of the binaries in this sample peaks at $13^{+14}_{-9}\mathrm{\,AU}$, which is consistent with previous measurements of early-M binaries. Hence, early-M binaries seem to occur in---on average---tighter systems than G binaries. At the same time they are significantly wider than field VLM binary stars. The distribution of mass ratios $q$ of primary and secondary stars was found to show an intermediate distribution between the strongly $q\rightarrow1$ peaked distribution of field VLM systems and the almost flat distribution of earlier-type stars. Consequently, we show evidence for relatively young, early-M binaries representing a transition between the well known earlier star distributions and the recently examined field VLM population characteristics. Despite the fact that this survey was dedicated to the search for faint brown dwarf and planetary mass companions, all planetary mass candidates were background objects. We exclude the existence of physical companions with masses greater than 10 Jupiter masses ($\mathrm{M_{Jup}}$) at separations of $\gtrsim 40\mathrm{\,AU}$ and masses greater than $24\mathrm{\,M_{Jup}}$ for separations $\gtrsim 10\mathrm{\,AU}$ around 37 of the 41 observed objects.
\end{abstract}

\keywords{binaries: general --- instrumentation: adaptive optics --- stars: late-type --- stars: low-mass, brown dwarfs}

\section {Introduction}
Binary and multiple stars have long provided a highly effective method of testing stellar evolution theory. As coeval systems, the intrinsic properties deduced for individual components (luminosity, effective temperature, color) should be consistent with models drawn from the same isochrone. Alternatively, parameters that are observationally well determined for one component (such as distance or metallicity) can be associated with the other component(s). This is particularly useful in characterizing the intrinsic properties of faint companions, including both low-mass stars (e.g. Gould \& Chanam{\'e} 2004) and brown dwarfs (e.g. Kirkpatrick et al. 2001)---of F, G, and early-K main sequence stars or subdwarfs. 

There is a long astronomical tradition of inverting the latter technique, and searching the environs of nearby stars for low luminosity companions. Effectively, the main-sequence stars act as cosmic lamp posts, each illuminating a small corner of the solar neighborhood. The efficiency of this search process is well illustrated by its success in identifying the archetypical ultracool M dwarf, VB\,10 \citep{bie44}, the first L dwarf, GD\,165B \citep{bec88}, and the first T dwarf, Gl\,229B \citep{nak95}. Subsequent studies have built on this foundation, using a variety of methods, including coronagraphy (e.g. Oppenheimer et al. 2001), wide-field ground-based imaging (e.g. Simons et al. 1996), and high-resolution imaging with both the {\it Hubble Space Telescope} ({\it HST}; e.g. Golimowski et al. 2004) and ground-based adaptive optics (AO) systems (e.g. Close et al. 2003; Siegler et al. 2005; Burgasser et al. 2006 and references within).

This paper describes our use of AO techniques on the Gemini telescope to search for low-mass brown dwarf and planetary mass companions around young, early-M dwarfs in the vicinity of the Sun. Section \ref{sampsel} describes the sample selection and section \ref{obs} describes our observations. Sections \ref{red} and \ref{ana} outline reduction and derived system parameters and section \ref{dis} discusses their implications.

\section {Sample selection}\label{sampsel}
The primary goal of our observing program is to identify very low-mass brown dwarfs as companions to nearby stars. That goal constrains our choice of targets. The probability of detecting a companion depends on the contrast with respect to the underlying flux distribution of the primary. The contrast, in turn, depends on the relative luminosity of primary and secondary, the angular separation, and the point-spread function (PSF). 

Lacking a long-lived central energy source, brown dwarfs cool and fade over relatively rapid timescales; moreover, low-mass brown dwarfs cool faster than high-mass brown dwarfs ($L \propto \tau^{-1.3} M^{2.64}$; Burrows et al. 2001). Thus, all other factors being equal, targeting young stars maximizes the probability of detecting low-mass brown dwarf companions. The nearest young star forming regions, however, lie at distances of more than $\sim\!150$ parsecs, while even the 10 to 20 Myr-old members of field associations, such as TW Hya and Tucana, are at distances of 50 to 100 parsecs from the Sun \citep{zuc04a}. Thus, the advantages offered by youth need to be balanced against the concomitant disadvantages of loss of linear resolution and flux.

Optimizing the contrast between primary and secondary presents a similar dilemma. In principle, very low-luminosity primaries offer the best prospects of detecting even lower luminosity secondaries; hence, (young) ultracool dwarfs (spectral types later than M7) might appear to be the best targets for our program. However, recent surveys have shown that the binary frequency decreases towards later spectral types, primarily because the maximum separation is a strong function of the total system mass. Few ultracool binaries are found in the field with separations exceeding 10 AU (Reid et al. 2001; Close et al. 2003; Burgasser et al. 2003, 2006; Bouy et al. 2003; Gizis et al. 2003; Mart{\'{\i}}n et al. 2003; Siegler et al. 2005), and it is helpful for any survey for very low-mass companions to focus on systems within the immediate Solar Neighborhood. The space density of ultracool dwarfs is relatively low, with only $\sim\!80$ L dwarfs within 20 parsecs of the Sun \citep{cru03}. Most of those systems are likely to be older than $\sim\!500$\,Myr \citep{all05}. 

Given these competing factors, we have chosen to target early-M dwarfs in the present survey. While these systems are more luminous than L dwarfs ($6 < M_\mathrm{J} < 11$ as compared with $11 < M_\mathrm{J} < 15$), they are much more numerous ($\sim\!2500$ systems within 20 parsecs) and support binary systems with separations as wide as 400\,AU. Moreover, the relative level of coronal activity provides a mechanism for selecting the youngest systems for observation in our program.

\begin{figure}
\includegraphics[angle=0,width=\columnwidth]{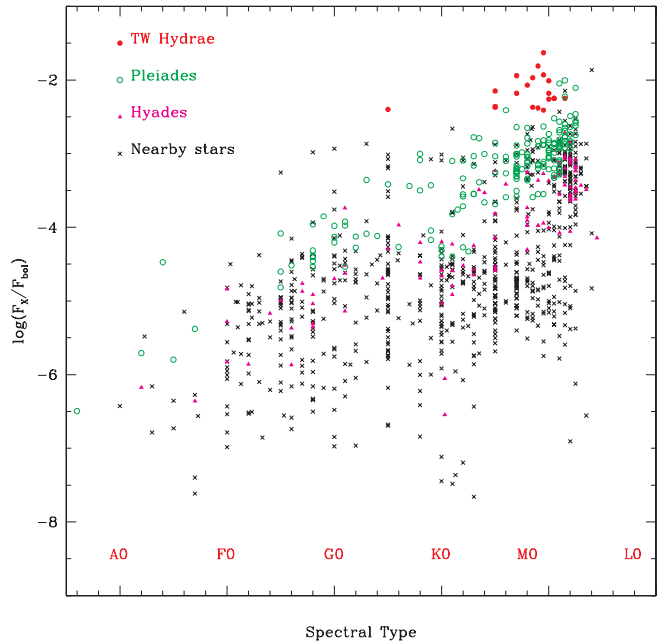}
\caption{X-ray activity as a function of age and spectral type. Data for the nearby stars are from \citet{hue99}, the Hyades data are taken from \citet{rei95b}, the Pleiades data from \citet{mic99}, and the TW Hydrae Association data from \citet{rei03a}.
\label{fig1}}
\end{figure}

M dwarfs have active chromospheres and coronae, characterized by optical emission (Ca II H\&K and Balmer lines) and X-ray emission, respectively. Figure \ref{fig1} shows X-ray activity (plotting $f_x/f_\mathrm{bol}$, where $f_x$ is the flux measured by ROSAT) as a function of spectral type and $\mathrm{V}-\mathrm{J}$ color for stars from the TW Hydrae association ($\sim\!10$ Myr; Reid 2003), the Pleiades ($\sim\!100$ Myr; Micela et al. 1999), the Hyades ($\sim\!600$ Myr; Reid et al. 1995b) and local field stars \citep{hue99}. There is a clear decrease in activity with age, with most of the field stars lying well below the cluster members. 

Using the data plotted in Figure \ref{fig1} as a guide, we have defined selection criteria in the [$\log (f_x/f_\mathrm{bol})$, $\mathrm{V}-\mathrm{J}$] plane that are designed to identify M dwarfs with activity levels comparable to or exceeding those of Pleiades members:
\begin {displaymath}
   \log{\frac{f_x}{f_\mathrm{bol}}} > -3.7 + 0.3(V-J), \quad (V-J) > 3.0\quad.
\end{displaymath}
Given the dispersion in activity levels at a given age, most single stars in this sample are likely to have ages younger than $\sim\!300$ Myr.

We have applied these criteria to two data sets: $\sim\!800$ proper motion stars from the 2MASS-based NStars survey described by \citet{rei04}; and $\sim\!380$ pre\-vi\-ous\-ly-un\-cata\-log\-ed late-type field dwarfs that show significant motion between the POSS\,I and 2MASS surveys (the Moving M sample, Reid, Cruz \& Allen, in prep.). Most of the former stars are drawn from the NLTT proper motion catalogue \citep{luy80}, but the sample also includes lower motion dwarfs from the Third Catalogue of Nearby Stars (CNS3; Gliese \& Jahreiss 1991). The effective proper motion limits of both samples correspond to $\sim\!0.2$ arcsec/yr, or a tangential velocity of 19 km/sec at 20\,pc.
It is important to note that this threshold does not bias against young stars. Systems like the Pleiades with a relative velocity of $\sim 29\mathrm{\,km/sec}$ with respect to the Sun \citep{jon96} or TW Hydrae with a relative motion of $\sim 23\mathrm{\,km/sec}$ with respect to the Sun \citep{rei03a} offer exemplary proof that young stellar objects do not need to have velocities within a few km/sec of the local standard of rest.
Distance estimates are based primarily on photometric or spectroscopic parallaxes. Both samples are drawn from the area covered by the 2MASS Second Incremental release ($\sim\!44\,\%$ of the sky), and both were cross-referenced against the {\it ROSAT} bright and faint source catalogues (Voges et al. 1999; 2000), searching for counterparts within 15 arcseconds. 196 stars from the first sample and 81 from the second are matched with ROSAT sources.

\begin{figure}
\begin{center}
\includegraphics[angle=0,width=\columnwidth]{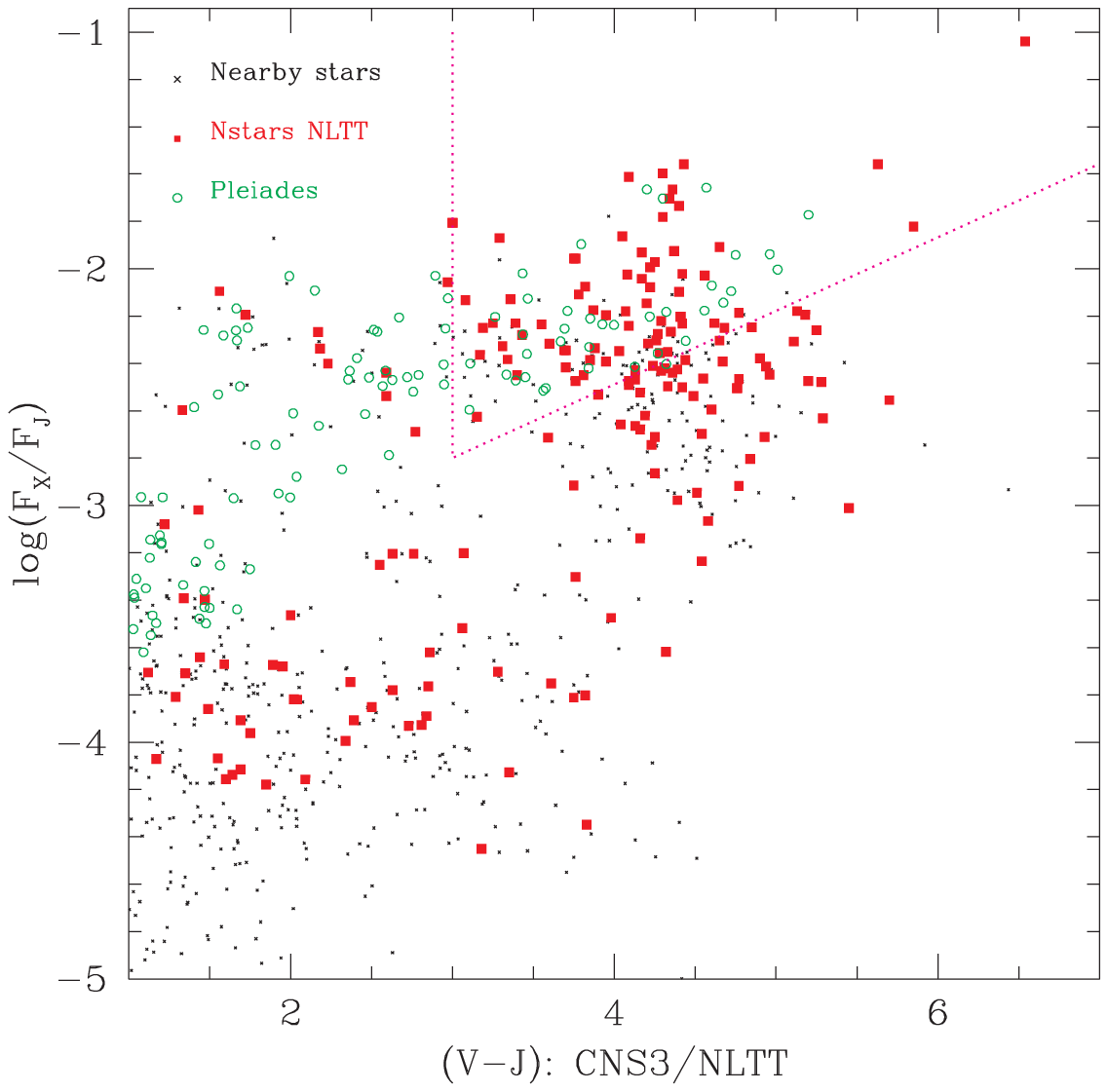}
\includegraphics[angle=0,width=\columnwidth]{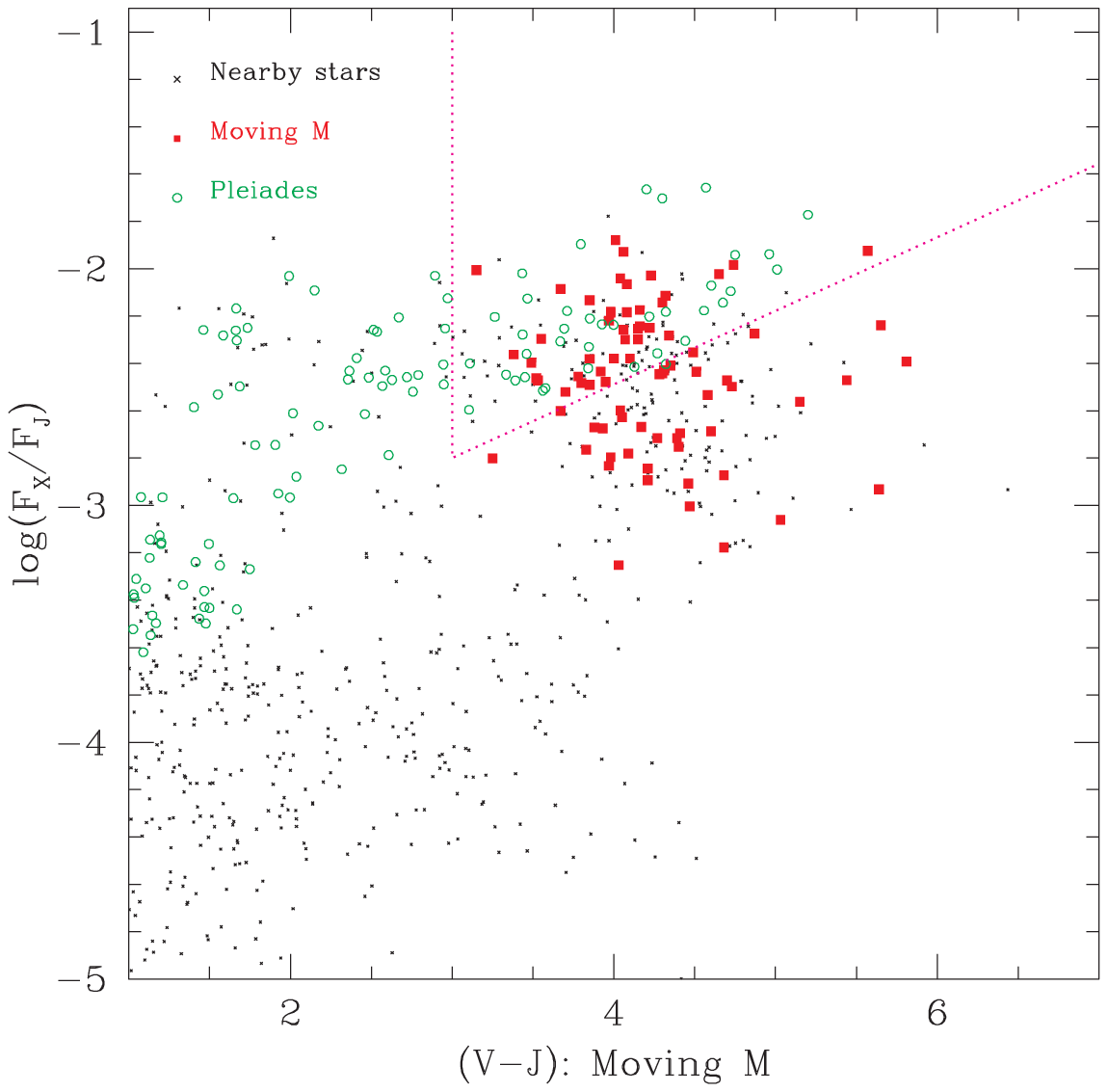}
\caption{Targets for this program were chosen based on the level of X-ray activity. The upper panel plots $\log(F_\mathrm{X}/F_\mathrm{bol})$ as a function of ($\mathrm{V}-\mathrm{J}$) color for the $\sim\!200$ ROSAT-detected sources from the NStars $d<20$ pc NLTT sample \citep{rei04}. The overall distribution is consistent with the larger \citet{hue99} nearby star sample. We have superposed X-ray activity data for members of the 100-Myr old Pleiades cluster, and the dotted lines mark the selection criteria that we have used to segregate candidate young stars in the NStars NLTT sample. The lower panel plots similar data for the newly-identified nearby star candidates in the ``Moving M'' sample (Reid, Cruz, \& Allen, in prep.).
\label{fig2}}
\end{center}
\end{figure}

Figure \ref{fig2} compares the coronal activity of the stars in these two samples against our selection criteria. A total of 92 stars exceed the Pleiades-like (100 Myr) activity threshold. However, 36 of these stars are known to be spectroscopic binaries, and it is likely that the high activity reflects rapid rotation in tidally-locked systems. We have excluded those stars from consideration in the current program, limiting observations to systems that are currently identified as single stars. Nonetheless, it is probable that some of the stars targeted in this program will eventually prove to be unrecognized spectroscopic binaries, rather than Pleiades-age single stars.

\section{Observations}\label{obs}
The current paper presents observations of a subset of the X-ray active stars from the NLTT/CNS3 and Moving M samples. All targeted objects are estimated to lie within 20 parsecs of the Sun.

All observations were taken in queue mode with the \mbox{8-m} Gemini North telescope using the facility Near-Infrared Imager (NIRI; Hodapp et al. 2003) in conjunction with the Altair AO system. The detector is a 1024$\times$1024 InSb Aladdin infrared hybrid array with 27$\micron$ pixels. The f/32 camera was used providing a $0\farcs022$/pixel platescale and 22$\arcsec\times22\arcsec$ field of view. The observations were conducted with the Cassegrain rotator purposefully disabled, leading to a fixed image pupil relative to camera and detector. 

For all observed targets, both short ($<\!10\mathrm{\,s}$) and deep (30\,s) exposures in the H and K$_\mathrm{s}$ filter bands were obtained. Each of the targets were imaged in a 5-position, $7\arcsec$-offset dithering pattern; 3 images were taken per dither position. The short frames were designed to primarily obtain unsaturated target images so as to reveal any near-equal mass companions at separations $\gtrsim\!0\farcs08$. They also served to provide astrometry and PSF sources. In a few cases even the minimum integration time of $0.2\,\mathrm{s}$ allowed by the NIRI IR camera led to saturated short frames due to the brightness of the target star. In these cases, other unsaturated sources were used for PSF fitting. All of the discovered physical binary systems were detected in the short frames.

The deep frames were designed to saturate the target star to gain sensitivity to potential very faint companions ($\Delta$H $>5\,$mag) at radial distances $\gtrsim\!1\arcsec$. Integrated counts were typically $\sim\!\,200$--$500\%$ in excess of full well depth at both H and K$_\mathrm{s}$ in the primary's core. Through the use of coadds, 30\,s integration times were obtained per image; 3 images per dither position gave 7.5\,min of total integration time per target.

Forty-one of the 56 objects from our survey sample were observed with the Gemini North telescope during semesters 2005A, 2005B, and 2006A. Table \ref{tab1} lists the 23 objects observed with no likely physical companions according to our sensitivity limits (examination of the detection limits is given in \S\ref{sen}). Four of the observed objects showed faint companions of currently undetermined nature. Follow-up observations are necessary to reveal whether the detections are consistent with proper motion objects and therefore physical companions. One additional object is believed to be a tight equal-mass binary system with an estimated separation comparable to the FWHM of $0\farcs08$ of these observations. Poor image quality probably due to bad weather, however, makes follow-up observations necessary to confirm its nature. None of the other binary images suffered from similar aberrations. Table \ref{tab2} shows the observable properties of the 12 binary systems observed, 7 of which are newly discovered binaries, as well as the parameters of the newly discovered triple system.

\begin{deluxetable}{lllcc}
\tabletypesize{\scriptsize}
\tablecaption{Young M0.5--M5.0 Stars Observed with No Physical Companion Detections Within the Sensitivity Limits\tablenotemark{a}\label{tab1}}
\tablewidth{\columnwidth}
\tablehead{
\colhead{2MASS Name} &
\colhead{Other Name} &
\colhead{K$_s$} &
\colhead{SpT} &
\colhead{Ref.} 
}
\startdata
J00115302+2259047 & LP\,348-40 & 8.00 & M3.5 & 1 \\
J01351393--0712517 &  & 8.08 & M4.0 & 2 \\ 
J01531133--2105433 & LP\,828-89 & 7.14 & M1.0 & 1 \\
J02085359+4926565 & G\,173-39 & 7.58 & M3.5 & 3 \\ 
J03472333--0158195 & HIP\,17695 & 6.93 & M3.0 & 2 \\ 
J03494325+2419046 & II\,Tau & 8.94 & M4.0 & 1 \\
J04244260--0647313 &  & 8.63 & M4.5 & 2 \\
J04373746--0229282 & Steph\,497 & 6.41 & M0.5 & 1 \\
J08081359+2106094 & LHS 5134 & 6.52 & M2.5 & 1 \\
J10193634+1952122 & Gl\,388 & 4.59 & M3.0 & 3 \\ 
J11032125+1337571 & LP\,491-51 & 7.91 & M3.5 & 1 \\
J11115176+3332111 & G\,119-62 & 7.50 & M3.5 & 3 \\ 
J14130492--1201262 & Gl\,540.2 & 8.16 & M4.5 & 3 \\
J14200478+3903014 & Steph\,1145 & 7.71 & M2.5 & 1 \\
J15215291+2058394 & Wo\,9520 & 5.76 & M1.5 & 1 \\
J15530484+4457446 &  & 9.25 & M3.5 & 4 \\
J18021660+6415445 & LP\,71-82 & 7.65 & M4.5 & 5 \\
J18130657+2601519 & LP\,390-16 & 8.07 & M4.0 & 1 \\ 
J19311257+3607300 & G\,125-15 & 8.83 & M4.5 & 1 \\
J20464360--1148132 & LP\,756-3 & 8.44 & M4.0 & 4 \\ 
J22143835--2141535 & BD-22\,5866 & 6.72 & M0.5 & 1 \\
J22464980+4420030 & Gl\,873 & 5.30 & M3.5 & 3 \\
J22515345+3145153 & Gl\,875.1 & 6.87 & M3.0 & 3 \\[-1.5ex]
\enddata
\tablenotetext{a}{No companions with mass $>$\,$10\mathrm{\,M_{Jup}}$ in more than $40\mathrm{\,AU}$ separation or $>$\,$24\mathrm{\,M_{Jup}}$ in more than $10\mathrm{\,AU}$ separation. Sensitivity limits are examined in section \S\ref{sen} and Figure \ref{fig8}.}
\tablerefs{(1) Reid et al. 2004; (2) Reid, Cruz, \& Allen (in prep.); (3) Reid et al. 1995a; (4) Riaz, Gizis \& Harvin 2006; (5) Reid et al. 2003}
\end{deluxetable}

\begin{deluxetable*}{llcr@{$\pm$}lr@{$\pm$}lr@{$\pm$}lr@{$\pm$}lcc}
\tabletypesize{\scriptsize}
\tablecaption{The Multiple Systems \label{tab2}}
\tablewidth{-479.16212pt}
\tablehead{
\colhead{2MASS Name} &
\colhead{Other Name} &
\colhead{SpT} &
\multicolumn{2}{c}{$\Delta H$} &
\multicolumn{2}{c}{$\Delta K_s$} &
\multicolumn{2}{c}{Sep. (arcsec)} &
\multicolumn{2}{c}{P.A. (deg)} &
\colhead{Date (UT)} &
\colhead{Ref.}
}
\startdata
J00285391+5022330 & G\,172-1 & M4.0\tablenotemark{1} & $0.50$ & $0.03$ & $0.49$ & $0.02$ & $0.426$ & $0.002$ & $32.13$ & $0.07$ & 05\,Aug\,05 &  TP \\
J01242767--3355086  & G\,274-24 & M5.0\tablenotemark{1} & $0.20$ & $0.03$ & $0.05$ & $0.02$ & $2.065$ & $0.002$ & $ 44.80$ & $0.07$ & 24\,Aug\,05 & TP \\
  &  &  & \multicolumn{2}{c}{ } & \multicolumn{2}{c}{ } & $2.01$ & $0.07$ & $46.38$ & $1.31$ & Jul\,99 & 5 \\
J04381255+2813001 & G\,39-29 & M4.0\tablenotemark{1} & $0.39$ & $0.03$ & $0.37$ & $0.03$ & $0.783$ & $0.002$ & $300.58$ & $0.09$ & 15\,Oct\,05 & TP \\
  &  &  & \multicolumn{2}{c}{ } & \multicolumn{2}{c}{ } & \multicolumn{2}{c}{$0.54$} & \multicolumn{2}{c}{$299$} & 13\,Sep\,02 &  6 \\
J05254166--0909123 & LP\,717-36 & M3.5\tablenotemark{2} & $0.46$ & $0.02$ & $0.42$ & $0.03$ & $0.537$ & $0.002$ & $69.40$ & $0.11$ & 14\,Oct\,05 & TP \\
J07285137--3014490 & GJ\,2060 & M0.5\tablenotemark{2} & $0.44$ & $0.30$\tablenotemark{a} & $0.44$ & $0.42$\tablenotemark{a} & $0.175$ & $0.011$ & $143.71$ & $1.54$ & 30\,Oct\,02 &  TP \\
  &  &  & \multicolumn{2}{c}{ } & \multicolumn{2}{c}{ } & \multicolumn{2}{c}{$0.3 ?$} & \multicolumn{2}{c}{?} & ? & 7 \\
J10364483+1521394 &  & M4.0\tablenotemark{3} & $1.15$ & $0.02$\tablenotemark{b} & $1.12$ & $0.01$\tablenotemark{b} & $1.061$ & $0.002$\tablenotemark{b} & $181.22$ & $0.07$\tablenotemark{b} & 18\,May\,06 &  TP \\
  &  &  & $0.05$ & $0.02$\tablenotemark{c} & $0.03$ & $0.02$\tablenotemark{c} & $0.189$ & $0.002$\tablenotemark{c} & $310.62$ & $0.14$\tablenotemark{c} &  &  \\
J13271966--3110394  & LHS\,2739 & M3.5\tablenotemark{1} & $0.14$ & $0.04$ & $0.17$ & $0.05$ & $0.544$ & $0.004$ & $333.79$ & $0.09$ & 12\,Jun\,05 &  TP \\
J13345147+3746195 &  & M3.5\tablenotemark{4} & $0.38$ & $0.03$ & $0.25$ & $0.03$ & $0.082$ & $0.003$ & $197.90$ & $0.27$ & 26\,Mar\,06 &  TP \\
J15553178+3512028 & G\,180-11 & M4.5\tablenotemark{1} & $1.99$ & $0.02$ & $1.97$ & $0.01$ & $1.571$ & $0.002$ & $261.23$ & $0.07$ & 27\,May\,05 &  TP \\
  &  &  & \multicolumn{2}{c}{ } & \multicolumn{2}{c}{ } & \multicolumn{2}{c}{$1.5$} & \multicolumn{2}{c}{$266$} & Mar\,98 & 8 \\
J17035283+3211456  & LP\,331-57 & M2.5\tablenotemark{1} & $1.53$ & $0.06$ & $1.53$ & $0.04$ & $1.130$ & $0.002$ & $136.53$ & $0.11$ & 22\,Apr\,05 & TP \\
J22332264--0936537  & Steph\,2018 & M3.0\tablenotemark{2} & $0.09$ & $0.04$ & $0.03$ & $0.03$ & $1.571$ & $0.003$ & $279.73$ & $0.14$ & 09\,Jun\,05 & TP \\
  &  &  & \multicolumn{2}{c}{ } & \multicolumn{2}{c}{ } & \multicolumn{2}{c}{$1.66$} & \multicolumn{2}{c}{$272.25$} & Jul\,97 & 8 \\
J23205766--0147373 & LP\,642-48 & M4.0\tablenotemark{4} & $0.32$ & $0.02$ & $0.27$ & $0.03$ & $0.099$ & $0.002$ & $325.06$ & $0.09$ & 12\,Jun\,05 & TP \\
J23581366--1724338 & LP\,764-40 & M2.0\tablenotemark{4} & $0.02$ & $0.02$ & $0.02$  & $0.02$ & $1.904$ & $0.002$ & $265.30$ & $0.88$ & 16\,Jul\,05 & TP\\[-1.5ex]
\enddata%
\tablenotetext{a}{$\mathrm{\Delta H}$ was estimated from artificial ghost, $\mathrm{\Delta K_s} = \mathrm{\Delta H}$ was assumed (see \S\ref{red} for more detail).}
\tablenotetext{b}{Numbers refer to the system consisting of A and B component of the triple system 2MASS\,J10364483+1521394.}
\tablenotetext{c}{Numbers refer to the system consisting of B and C component of 2MASS\,J10364483+1521394.}
\tablerefs{(TP) This paper; (1) Reid et al. 1995a; (2) Reid et al. 2003; (3) Reid, Cruz, \& Allen (in prep.); (4) Riaz, Gizis \& Harvin 2006; (5) Jao et al. 2003; (6) Beuzit et al. 2004; (7) Zuckerman et al. 2004; (8) McCarthy et al. 2001}
\end{deluxetable*}

\section{Reduction}\label{red}
The data reduction used an AO data reduction pipeline written in the IRAF language as described in \citet{clo02}. The pipeline outputs final unsaturated exposures in $\mathrm{H}$ and $\mathrm{K_s}$ band ($\mathrm{FWHM}\sim 0\farcs08$), and deep $450\mathrm{\,s}$ ($15\times30\mathrm{\,s}$) exposures for each observed object. The dithering produces a central $10\arcsec\times10\arcsec$ high signal-to-noise-ratio (S/N) region in a $30\arcsec\times30\arcsec$ final image centered on the object.

\begin{figure*}[tb]
\includegraphics[angle=0,width=\textwidth]{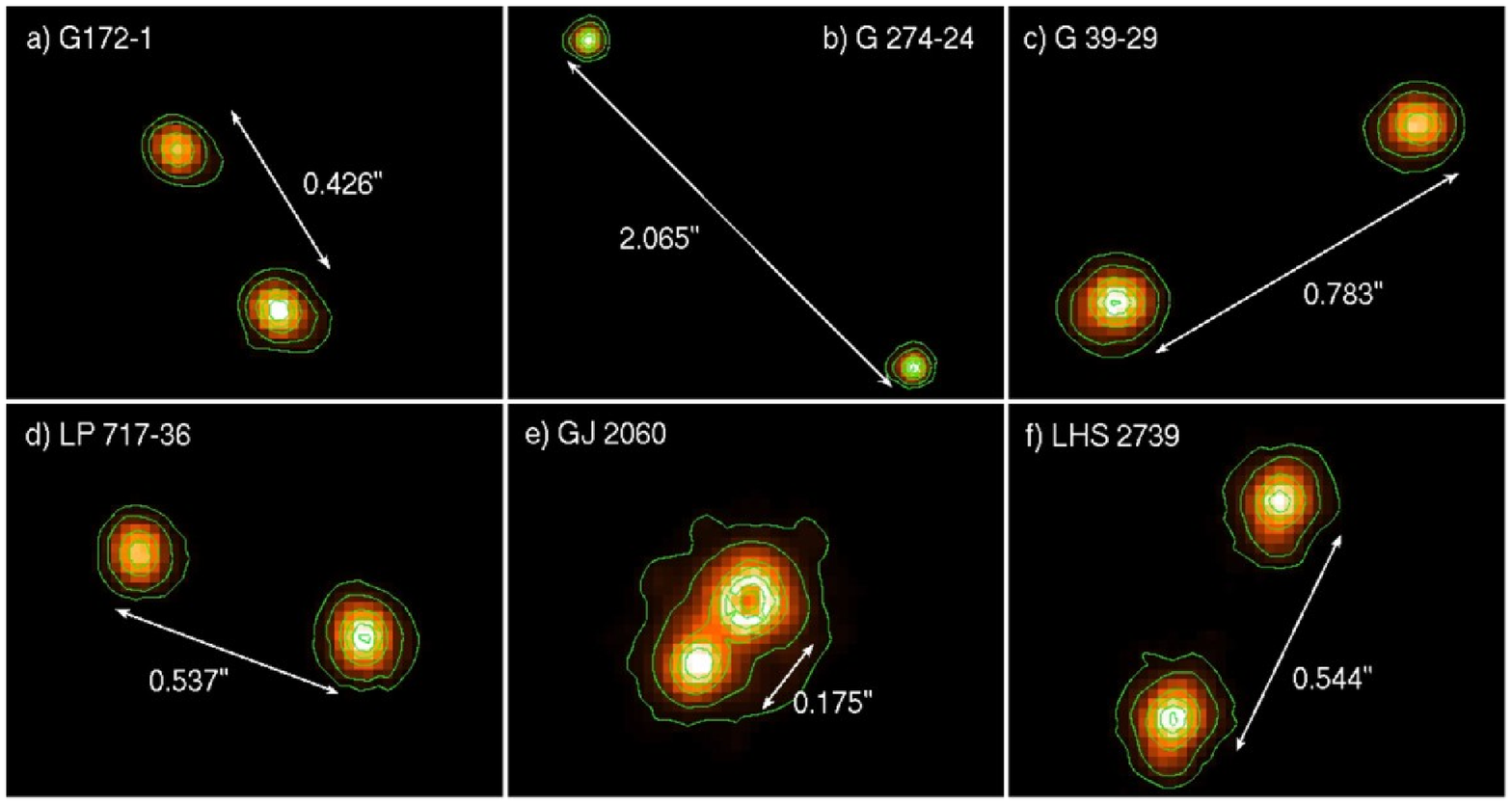}
\caption{The 0.2\,s $\mathrm{K_s}$ images of the observed binaries. The platescale is 0.022\,arcsec/pixel, the contours are linear at the 90\%, 70\%, 50\%, 30\%, 10\%, and 5\% levels. North is up and east is to the left. The dark central region in the A component of GJ 2060 is due to source saturation.
\label{fig3}} 
\end{figure*}

\begin{figure*}[tb]
\includegraphics[angle=0,width=\textwidth]{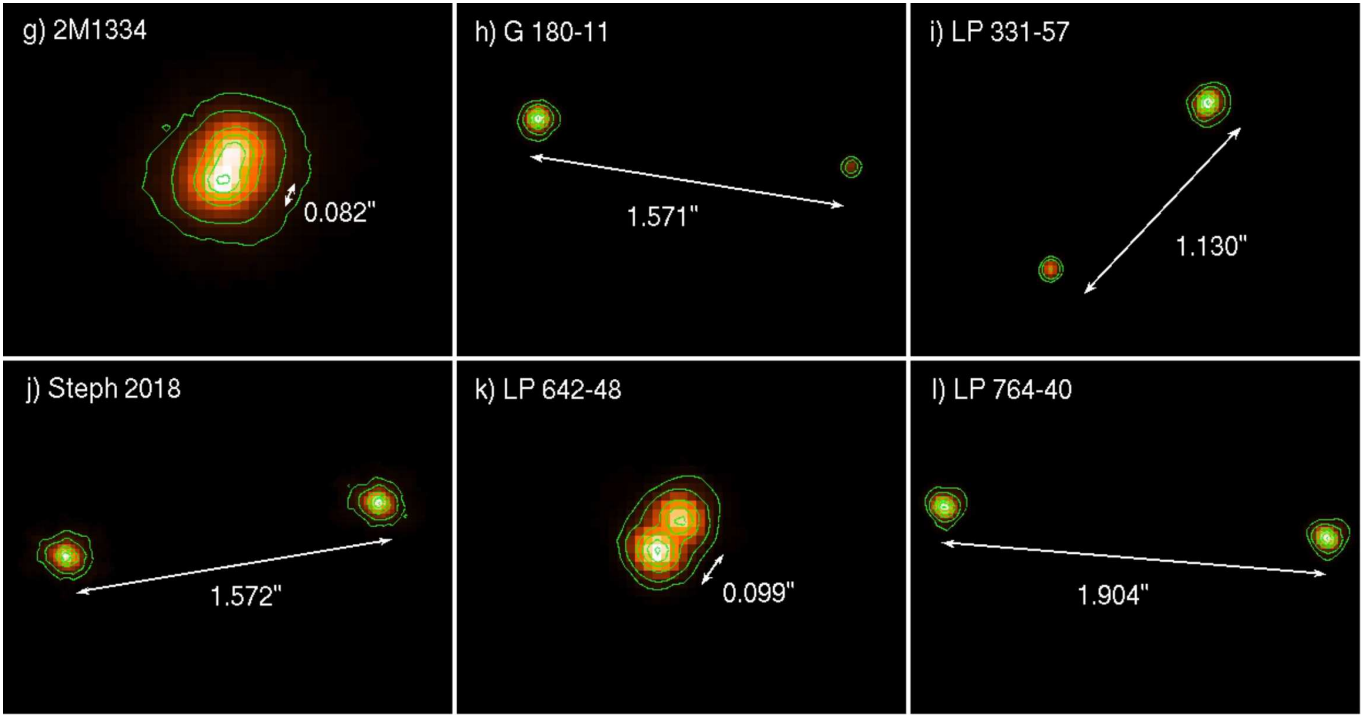}
\caption{Same as Figure \ref{fig3}.
\label{fig4}} 
\end{figure*}

\begin{figure}[tb]
\includegraphics[angle=0,width=\columnwidth]{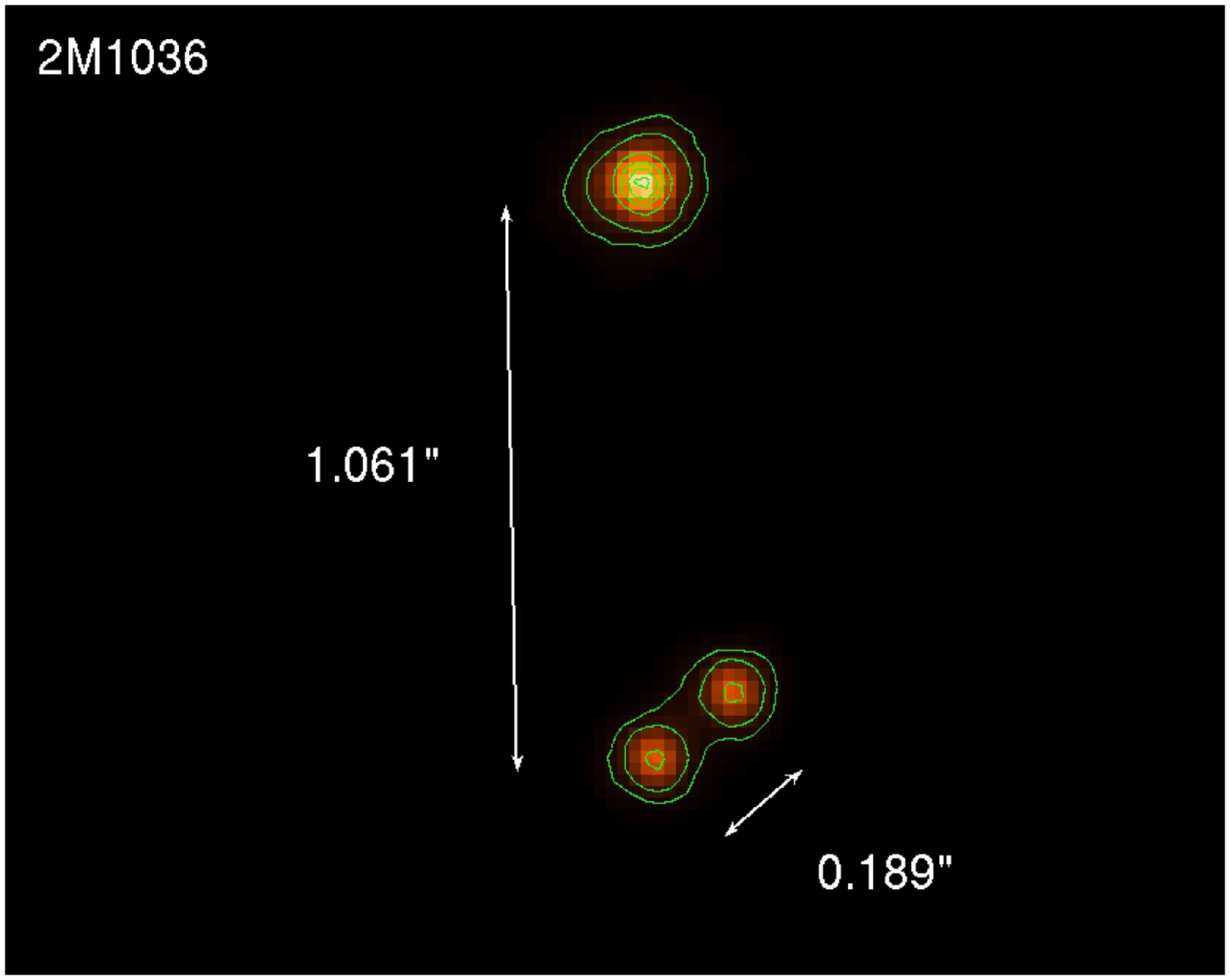}
\caption{Triple system 2MASS\,J10364483+1521394. Image parameters and contours are as in Figure \ref{fig3}.
\label{fig5}} 
\end{figure}

Since the Cassegrain rotator was disabled, the residual PSF aberrations including telescope primary and secondary structures remained static and fixed in the final images, allowing for comparative identification of faint companions without confusion from rotating PSF artifacts. The pipeline compensates for this by rotating each image by the parallactic angle before combining as described in \citet{clo03}. In order to detect faint companions within $1\arcsec$ distance to the central star, we filtered out the low spatial frequency components of the deep images leaving behind only the high frequency residuals in the PSF (unsharp masking). Figures \ref{fig3} and \ref{fig4} show images of the discovered binaries; Figure \ref{fig5} shows the triple system.

Photometry was performed using the ALLSTAR PSF-fitting task in IRAF. PSF stars were selected from similar magnitude targets observed either the same night (if available) or as close as possible to the observation date, at similar airmass. Uncertainties in $\Delta\mathrm{mag}$ and separations listed in Table \ref{tab2} were determined by comparing the results employing two distinct PSF stars for each binary. In the case of GJ\,2060, however, saturated raw images ruled out the use of ALLSTAR. Instead, we used an artificial ghost that appeared in every bright image in the H filter at fixed separation and orientation relative to the detector. This ghost, probably due to secondary reflection in the H filter, remained unsaturated in the long exposures and responded linearly to the flux of the generating object. By averaging the ratio of ghost flux difference to binary flux difference of two similar binaries (LP 717-36 and G 39-29) we were able to estimate $\Delta\mathrm{H}$ for GJ\,2060. Since there is no ghost in the $\mathrm{K_s}$ images, $\Delta\mathrm{K_s}=\Delta\mathrm{H}$ was assumed which is reasonable given that on average $\Delta\mathrm{H}-\Delta\mathrm{K_s}$ for the other binaries is consistent with 0. To account for uncertainties in this case, a larger error was assumed. The binary data are listed in Table \nolinebreak \ref{tab2}.

\subsection{Spectral Types, Absolute Magnitudes, \& Distances}\label{spt}
\begin{figure}[tb]
\includegraphics[angle=90,width=\columnwidth]{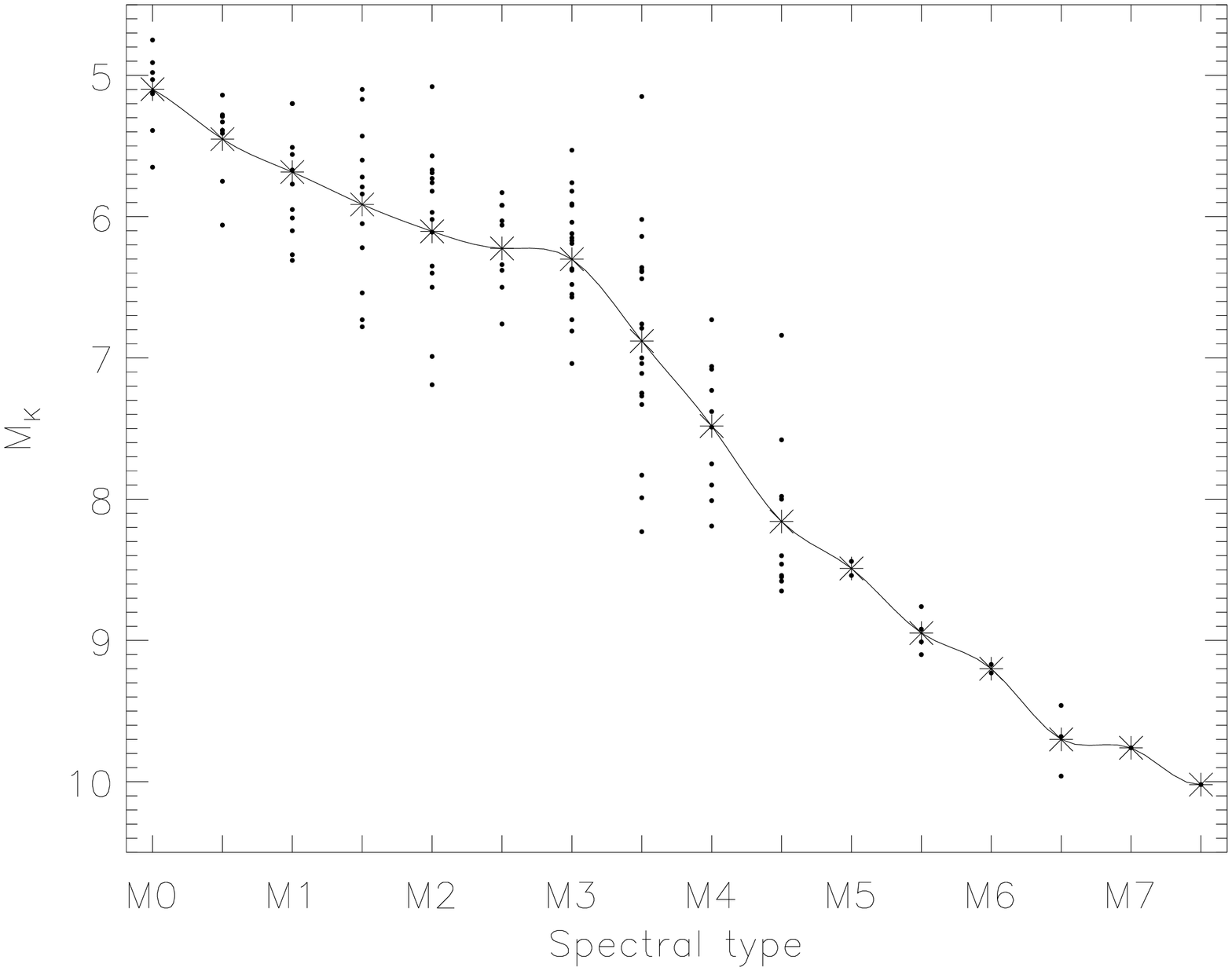}
\caption{Absolute K magnitude vs. spectral type for M stars, compiled from visual K band data and distances from \citet{kir96}, \citet{rei95b}, and \citet{haw96}. The dots show all objects with distance uncertainties of less than 1\,pc. The asterisks denote the average over these M$_\mathrm{K}$ for each spectral type. The line shows the spline interpolation of the averages.
\label{fig6}} 
\end{figure}

We do not have spatially resolved spectra of any of the new multiple systems. Moreover, intrinsic scatter in the $\mathrm{H}-\mathrm{K_s}$ colors of M dwarfs do not allow us to derive accurate spectral types from our observed $(\mathrm{H}-\mathrm{K_s})_\mathrm{A}$ and $(\mathrm{H}-\mathrm{K_s})_\mathrm{B}$ colors. However, we can estimate component spectral types and absolute magnitudes using averaged absolute $\mathrm{M_K}$ magnitudes for each spectral type [see Figure \ref{fig6} for the function $\mathrm{M_K}(\mathrm{SpT})$]. Component spectral types were estimated by assuming a flux-weighted relation between component and integrated spectral types of $\mathrm{SpT_{int}} = (f_\mathrm{A} \mathrm{SpT_\mathrm{A}} + f_\mathrm{B} \mathrm{SpT_\mathrm{B}})/(f_\mathrm{A}+f_\mathrm{B})$, where $f_\mathrm{A}$ and $f_\mathrm{B}$ are flux values derived from $\mathrm{K_s}$ for primary and secondary component respectively. We have evidence that a linear relation can be used to estimate SpT$_\mathrm{A}$ and SpT$_\mathrm{B}$ from SpT$_\mathrm{int}$ since the TiO and VO band strengths from which the spectral types are derived are quite linear for M0--M7 \citep{cru02}. The above relation defines possible pairs of spectral types $(\mathrm{SpT_\mathrm{A}},\mathrm{SpT_\mathrm{B}})$ for each binary. Another restriction on spectral types is given by $\mathrm{M_K}(\mathrm{SpT})$ in combination with our measured values for $\Delta\mathrm{K_s}$: $\Delta\mathrm{K_s}=\mathrm{M_K}(\mathrm{SpT_B})-\mathrm{M_K}(\mathrm{SpT_A})$ defines another relation between SpT$_\mathrm{A}$ and SpT$_\mathrm{B}$. Intersecting these two relations we find an unambiguous combination of $\mathrm{SpT_\mathrm{A}}$ and $\mathrm{SpT_\mathrm{B}}$ for each binary. Typically $\Delta$K is small---for example G\,172-1 has $\Delta\mathrm{K}=0.49$ and SpT$_\mathrm{int}=\mathrm{M}4$---and so we find SpT$_\mathrm{A}\approx\mathrm{M}4$ and SpT$_\mathrm{B}\approx\mathrm{M}4.5$. However,  in the case of G\,180-11 $\Delta\mathrm{K}=1.97$; therefore, we estimate SpT$_\mathrm{A}=\mathrm{M}4\pm0.5$ and SpT$_\mathrm{B}=\mathrm{M}7\pm2.0$ from SpT$_\mathrm{int}=\mathrm{M}4.5$. Equivalent considerations were made to derive parameter values for the A, B, and C components of the triple system 2M1036, exploiting the fact that the B and C component are of equal magnitude.

Absolute $\mathrm{{M_K}_s}$ magnitudes were derived from applying $\mathrm{M_K}(\mathrm{SpT})$ on interpolated spectral type values. We account for the conversion from K to K$_\mathrm{s}$ band for reasons of completeness, although the difference turns out to be negligible for most of the binaries. Using a sample of stars with published values for K and K$_\mathrm{s}$ magnitudes, differences between magnitudes in these two filter bands have been found to scatter around 0, with an amplitude of $\sim\!0.05\mathrm{\,mag}$. The uncertainties of the individual M$_\mathrm{K_s}$ values account for this incertitude in conversion by adding in an extra uncertainty of 0.035. From apparent and absolute magnitudes distances were calculated. Derived component spectral types, absolute magnitudes, and distance estimates are listed in Table \ref{tab3}.

\begin{deluxetable*}{lr@{$\pm$}lr@{$\pm$}lr@{$\pm$}lr@{$\pm$}lr@{$\pm$}lr@{$^+_-$}lr@{$\pm$}lr@{$^+_-$}l}
\tabletypesize{\scriptsize}
\tablecaption{Summary of the New Multiples' Components\label{tab3}}
\tablewidth{-479.16212pt}
\tablehead{
\colhead{Name} &
\multicolumn{2}{c}{{$H$}} &
\multicolumn{2}{c}{{$K_s$}} &
\multicolumn{2}{c}{{SpT\tablenotemark{a}}}&
\multicolumn{2}{c}{{$M_{K_s}$\tablenotemark{b}}} &
\multicolumn{2}{c}{{$d_\mathrm{phot}$ (pc)}} &
\multicolumn{2}{c}{{Mass (M$_{\sun}$)\tablenotemark{c}}} &
\multicolumn{2}{c}{{Sep. (AU)}} &
\multicolumn{2}{c}{{P (yr)\tablenotemark{d}}} 
}
\startdata
G\,172-1A & $8.82$ & $0.03$ & $8.52$ & $0.02$ & M$\,4.0$ & $0.5$ & $7.33$ & $0.65$ & $17.4$ & $4.5$ & $0.24$ & $^{0.11}_{0.07}$ & $ 7.4$ & $2.2$ & $40$ & $^{19}_{18}$ \\
G\,172-1B & $9.32$ & $0.03$ & $9.02$ & $0.02$ & M$\,4.5$ & $0.5$ & $7.82$ & $0.57$ & \multicolumn{2}{c}{ } & $0.19$ & $^{0.07}_{0.05}$ & \multicolumn{2}{c}{ } & \multicolumn{2}{c}{ } \\
G\,274-24A & $9.31$ & $0.05$ & $8.97$ & $0.03$ & M$\,5.0$ & $0.5$ & $8.46$ & $0.40$ & $12.6$ & $2.3$ & $0.14$ & $^{0.03}_{0.02}$ & $26.0$ & $4.8$ & $330$ & $^{ 95}_{ 93}$ \\
G\,274-24C\tablenotemark{e} & $9.52$ & $0.05$ & $9.02$ & $0.03$ & M$\,5.0$ & $0.5$ & $8.52$ & $0.40$ & \multicolumn{2}{c}{ } & $0.13$ & $^{0.03}_{0.02}$ & \multicolumn{2}{c}{ } & \multicolumn{2}{c}{ } \\
G\,39-29A & $8.19$ & $0.02$ & $8.20$ & $0.02$ & M$\,4.0$ & $0.5$ & $7.39$ & $0.65$ & $14.5$ & $3.8$ & $0.23$ & $^{0.10}_{0.07}$ & $11.4$ & $3.4$ & $76$ & $^{36}_{35}$ \\
G\,39-29B & $8.58$ & $0.03$ & $8.57$ & $0.02$ & M$\,4.5$ & $0.5$ & $7.76$ & $0.57$ & \multicolumn{2}{c}{ } & $0.20$ & $^{0.07}_{0.05}$ & \multicolumn{2}{c}{ } & \multicolumn{2}{c}{ } \\
LP\,717-36A & $8.43$ & $0.03$ & $8.19$ & $0.03$ & M$\,3.5$ & $0.5$ & $6.66$ & $0.50$ & $20.2$ & $4.7$ & $0.36$ & $^{0.09}_{0.09}$ & $10.9$ & $2.5$ & $58$ & $^{21}_{21}$ \\
LP\,717-36B & $8.89$ & $0.03$ & $8.60$ & $0.03$ & M$\,4.0$ & $0.5$ & $7.06$ & $0.64$ & \multicolumn{2}{c}{ } & $0.28$ & $^{0.11}_{0.08}$ & \multicolumn{2}{c}{ } & \multicolumn{2}{c}{ } \\
GJ\,2060A\tablenotemark{f} & $6.52$ & $0.13$ & $6.27$ & $0.17$ & M$\,0.5$ & $0.5$ & $5.28$ & $0.27$ & $15.8$ & $2.3$ & $0.59$ & $^{0.05}_{0.05}$ & $ 2.8$ & $0.4$ & $ 6$ & $ ^{ 1}_{ 1}$ \\
GJ\,2060B\tablenotemark{f} & $6.96$ & $0.18$ & $6.71$ & $0.26$ & M$\,1.5$ & $1.0$ & $5.72$ & $0.47$ & \multicolumn{2}{c}{ } & $0.52$ & $^{0.08}_{0.08}$ & \multicolumn{2}{c}{ } & \multicolumn{2}{c}{ } \\
2M1036A & $8.71$ & $0.03$ & $8.48$ & $0.03$ & M$\,3.5$ & $0.5$ & $7.02$ & $0.62$ & $19.6$ & $4.6$ & $0.29$ & $^{0.11}_{0.08}$ & $20.8$ & $6.0$ & $157$ & $^{ 97}_{ 93}$\tablenotemark{g}\\
2M1036B & $9.86$ & $0.03$ & $9.60$ & $0.03$ & M$\,4.5$ & $0.5$ & $8.14$ & $0.50$ & \multicolumn{2}{c}{ } & $0.16$ & $^{0.05}_{0.04}$ & $3.7$ & $1.1$ & $16$ & $^{10}_{ 9}$\tablenotemark{g} \\
2M1036C & $9.91$ & $0.03$ & $9.60$ & $0.03$ & M$\,4.5$ & $0.5$ & $8.14$ & $0.50$ & \multicolumn{2}{c}{ } & $0.16$ & $^{0.05}_{0.04}$ & \multicolumn{2}{c}{ } & \multicolumn{2}{c}{ } \\
LHS\,2739A & $9.45$ & $0.06$ & $9.19$ & $0.03$ & M$\,3.5$ & $0.5$ & $6.79$ & $0.55$ & $30.2$ & $7.7$ & $0.33$ & $^{0.10}_{0.09}$ & $16.4$ & $4.2$ & $108$ & $^{43}_{43}$ \\
LHS\,2739B & $9.59$ & $0.06$ & $9.36$ & $0.04$ & M$\,3.5$ & $0.5$ & $6.96$ & $0.62$ & \multicolumn{2}{c}{ } & $0.30$ & $^{0.11}_{0.09}$ & \multicolumn{2}{c}{ } & \multicolumn{2}{c}{ } \\
2M1334A & $9.73$ & $0.03$ & $9.52$ & $0.02$ & M$\,3.5$ & $0.5$ & $6.75$ & $0.65$ & $35.8$ & $9.4$ & $0.34$ & $^{0.11}_{0.10}$ & $2.9$ & $0.9$ & $8$ & $^{4}_{4}$ \\
2M1334B & $10.10$ & $0.03$ & $9.77$ & $0.02$ & M$\,3.5$ & $0.5$ & $7.00$ & $0.57$ & \multicolumn{2}{c}{ } & $0.30$ & $^{0.10}_{0.08}$ & \multicolumn{2}{c}{ } & \multicolumn{2}{c}{ } \\
G\,180-11A & $8.42$ & $0.03$ & $8.20$ & $0.02$ & M$\,4.0$ & $0.5$ & $7.70$ & $0.62$ & $12.6$ & $3.6$ & $0.20$ & $^{0.08}_{0.06}$ & $19.8$ & $5.7$ & $216$ & $^{101}_{ 97}$ \\
G\,180-11B & $10.42$ & $0.04$ & $10.18$ & $0.03$ & M$\,7.0$ & $2.0$ & $9.68$ & $1.17$ & \multicolumn{2}{c}{ } & $0.08$ & $^{0.06}_{0.03}$ & \multicolumn{2}{c}{ } & \multicolumn{2}{c}{ } \\
LP\,331-57A & $7.54$ & $0.02$ & $7.32$ & $0.02$ & M$\,2.0$ & $0.5$ & $6.04$ & $0.14$ & $18.0$ & $1.2$ & $0.46$ & $^{0.02}_{0.02}$ & $20.3$ & $1.4$ & $143$ & $^{17}_{16}$ \\
LP\,331-57B & $9.06$ & $0.05$ & $8.85$ & $0.04$ & M$\,4.0$ & $0.5$ & $7.58$ & $0.64$ & \multicolumn{2}{c}{ } & $0.22$ & $^{0.09}_{0.06}$ & \multicolumn{2}{c}{ } & \multicolumn{2}{c}{ } \\
Steph\,2018A & $8.66$ & $0.03$ & $8.42$ & $0.03$ & M$\,3.0$ & $0.5$ & $6.29$ & $0.33$ & $26.7$ & $4.1$ & $0.42$ & $^{0.05}_{0.06}$ & $41.9$ & $6.4$ & $382$ & $^{89}_{90}$ \\
Steph\,2018B & $8.76$ & $0.03$ & $8.45$ & $0.03$ & M$\,3.0$ & $0.5$ & $6.32$ & $0.33$ & \multicolumn{2}{c}{ } & $0.42$ & $^{0.05}_{0.06}$ & \multicolumn{2}{c}{ } & \multicolumn{2}{c}{ } \\
LP\,642-48A & $9.39$ & $0.03$ & $9.12$ & $0.02$ & M$\,4.0$ & $0.5$ & $7.35$ & $0.64$ & $22.6$ & $6.7$  & $0.25$ & $^{0.18}_{0.11}$ & $2.2$ & $0.7$ & $6$ & $^{ 3}_{ 3}$ \\
LP\,642-48B & $9.72$ & $0.03$ & $9.39$ & $0.03$ & M$\,4.0$ & $0.5$ & $7.62$ & $0.64$ & \multicolumn{2}{c}{ } & $0.21$ & $^{0.17}_{0.09}$ & \multicolumn{2}{c}{ } & \multicolumn{2}{c}{ } \\
LP\,746-40A & $8.44$ & $0.03$ & $8.18$ & $0.02$ & M$\,2.0$ & $0.5$ & $6.10$ & $0.16$ & $26.1$ & $2.0$  & $0.45$ & $^{0.05}_{0.05}$ & $49.7$ & $ 3.8$ & $477$ & $^{ 58}_{ 58}$ \\
LP\,746-40B & $8.46$ & $0.03$ & $8.20$ & $0.03$ & M$\,2.0$ & $0.5$ & $6.11$ & $0.16$ & \multicolumn{2}{c}{ } & $0.45$ & $^{0.05}_{0.05}$ & \multicolumn{2}{c}{ } & \multicolumn{2}{c}{ }\\[-1ex] 
\enddata
\tablenotetext{a}{Spectral type estimated by using flux weighted relation $\mathrm{SpT_{int}} = (f_\mathrm{A} \mathrm{SpT_\mathrm{A}} + f_\mathrm{B} \mathrm{SpT_\mathrm{B}})/(f_\mathrm{A}+f_\mathrm{B})$ in combination with $\Delta \mathrm{K_s}$ values and an $\mathrm{M_K}$-SpT relation as described in \S\ref{spt} (valid for $\mathrm{M0}\lesssim\mathrm{SpT}\lesssim\mathrm{M7.5}$).}
\tablenotetext{b}{$\mathrm{{M_K}_s}$ values are based on spectral types by use of an absolute $\mathrm{M_K}$ magnitude vs. spectral type relation (see \S\ref{spt}). Conversion from K to K$_\mathrm{s}$ was accounted for by adding an uncertainty of 0.035\,mag to $\mathrm{{M_K}_s}$ values (\S \ref{spt}).}
\tablenotetext{c}{Masses are estimated using the evolutionary models of \citet{bar98} assuming an age of 300\,Myr.}
\tablenotetext{d}{Period estimates include a 1.26 multiplication of the projected separations compensating for random inclinations and eccentricities \citep{fis92}.}
\tablenotetext{e}{The companion to G\,274-24A is part of a triple system with a wide ($37.8\arcsec$) secondary component as reported in \citet{jao03}, who refer to the close-in companion as G\,274-24C. We adopt this notation.}
\tablenotetext{f}{Values were derived using the assumption that $\mathrm{\Delta K_s} = \mathrm{\Delta H}$ (see \S\ref{red} for more detail)}
\tablenotetext{g}{The periods of the triple 2M1036 are calculated as A around (BC) and B around C. For the calculation of the former system, the total mass of the B+C system was used.}
\end{deluxetable*}

$\mathrm{M_K}(\mathrm{SpT})$ is derived from main-sequence stars (5\,Gyr). Young ($\lesssim\!300$\,Myr), low-mass objects like those in this sample have not yet reached the main sequence and are therefore slightly brighter than estimated. According to the models of \citet{bar98}, this difference is negligible ($<0.05\mathrm{\,mag}$ in K$_\mathrm{s}$) for objects brighter than $\mathrm{M_K}\sim 8.9$. The companion to G\,180-11, however, is estimated to have a magnitude of $\mathrm{M_{K_s}}= 9.7\pm1.2$. Its mass estimate (see \S \ref{mass}) therefore suffers from an additional uncertainty due to this effect.

\subsection{Masses and Periods}\label{mass}
We estimate masses for each of the binary components from their M$_\mathrm{K_s}$ magnitudes using 300\,Myr evolutionary models from \citet{bar98}. \citet{del00} compared 5\,Gyr Baraffe models to observational data finding very good agreement in K band. Although the objects in this paper are believed to be much younger and there exists only little observational data constraining a mass luminosity relation for these low mass objects, we expect a sufficiently high accuracy for the 300\,Myr models. The inaccuracies of the models (see Close et al. 2006) are assumed to be small compared to our uncertainties in spectral type. From this the binary periods are estimated by applying Kepler's third law, $P=a^{3/2}/\sqrt{M}$, using separations scaled by a factor of 1.26 to account for random inclinations and eccentricities, as derived in \citet{fis92}.

The calculation of the period of the triple system is based on the assumption that components B and C orbit each other in a tight system and they in turn orbit the primary component A in the same plane. This is a more or less likely assumption, given that gravitationally bound systems consisting of three objects of comparable mass are normally stable in a hierarchical configuration like that assumed and a few pathological situations. Thus, we list two periods, one for the BC system and one for A(BC). However, we cannot exclude a non-relaxed system in a more complicated configuration or a different hierarchy [e.g. the proximity of B and C being due to projection and therefore C revolving around (AB) in a wide orbit]. Table \ref{tab3} lists estimated masses and periods for the binaries and the triple system.

\subsection{Sensitivity to the Detection of Faint Companions}\label{sen}
\begin{figure}[tb]
\includegraphics[angle=0,width=\columnwidth]{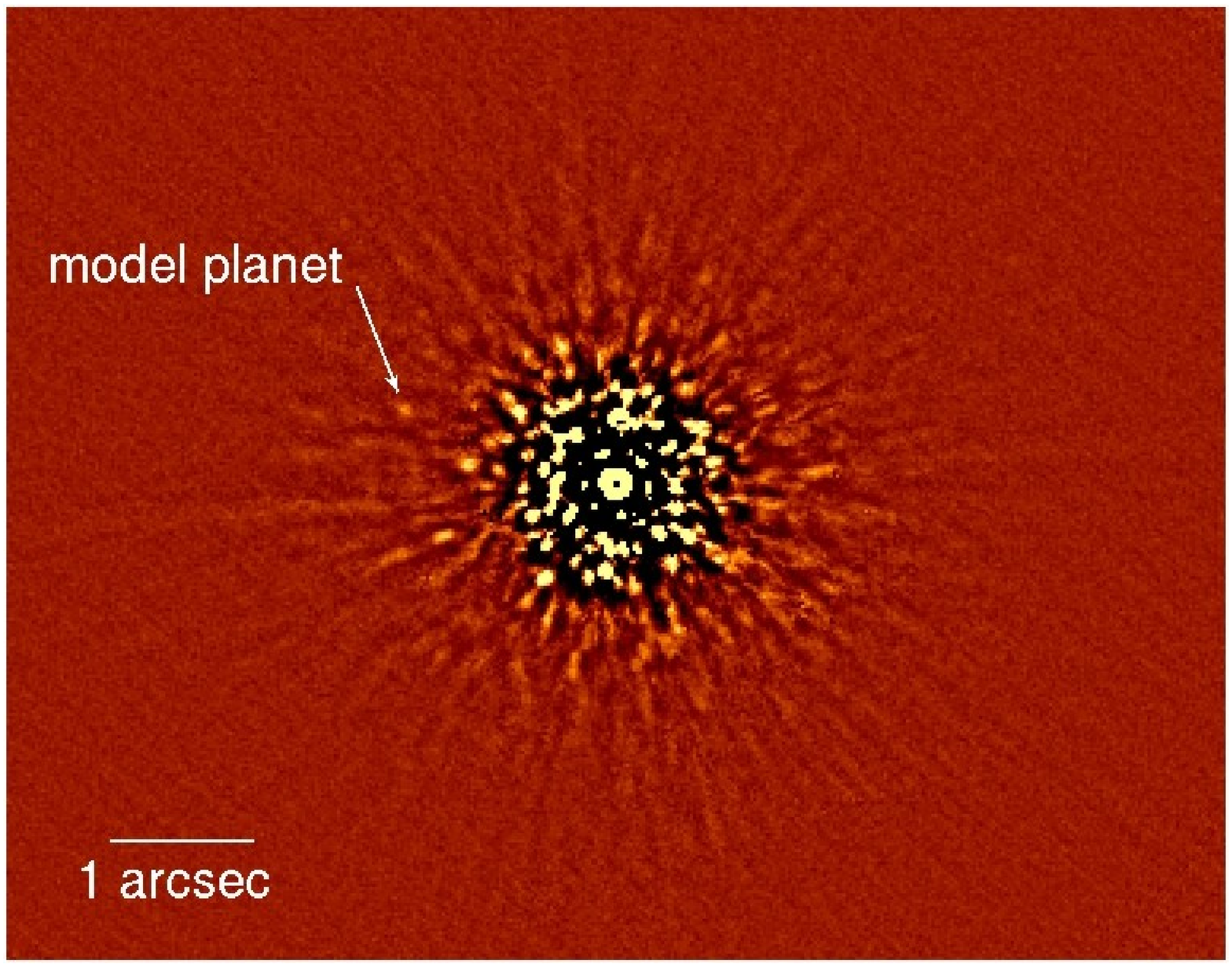}
\caption{Reduced $15\!\times\!30\mathrm{\,s}$ $\mathrm{K_s}$ exposure of LP\,756-3, unsharp masked (low spatial frequencies removed). A fake planet at the $5\sigma$ detection limit was inserted by scaling down the brightness of an unsaturated image of LP\,756-3 and adding it to the deep exposure. The detectable magnitude difference in $\mathrm{K_s}$-band is as high as $\Delta\mathrm{K_s}=10.9$ at a separation of $1\farcs5$. This corresponds to a $\sim\!10\mathrm{\,M_{Jup}}$ planet 23\,AU from the central star from the models of \citet{bar03} at 500\,Myr. The companion can be distinguished from speckle noise by ``blinking'' the image with an image of another star in the same wave band. Because the Cassegrain rotator was purposefully disabled, speckles would remain fixed in both images, while the companion reveals itself by appearing in only one of the images.
\label{fig7}} 
\end{figure}

\begin{figure}[tb]
\includegraphics[angle=90,width=\columnwidth]{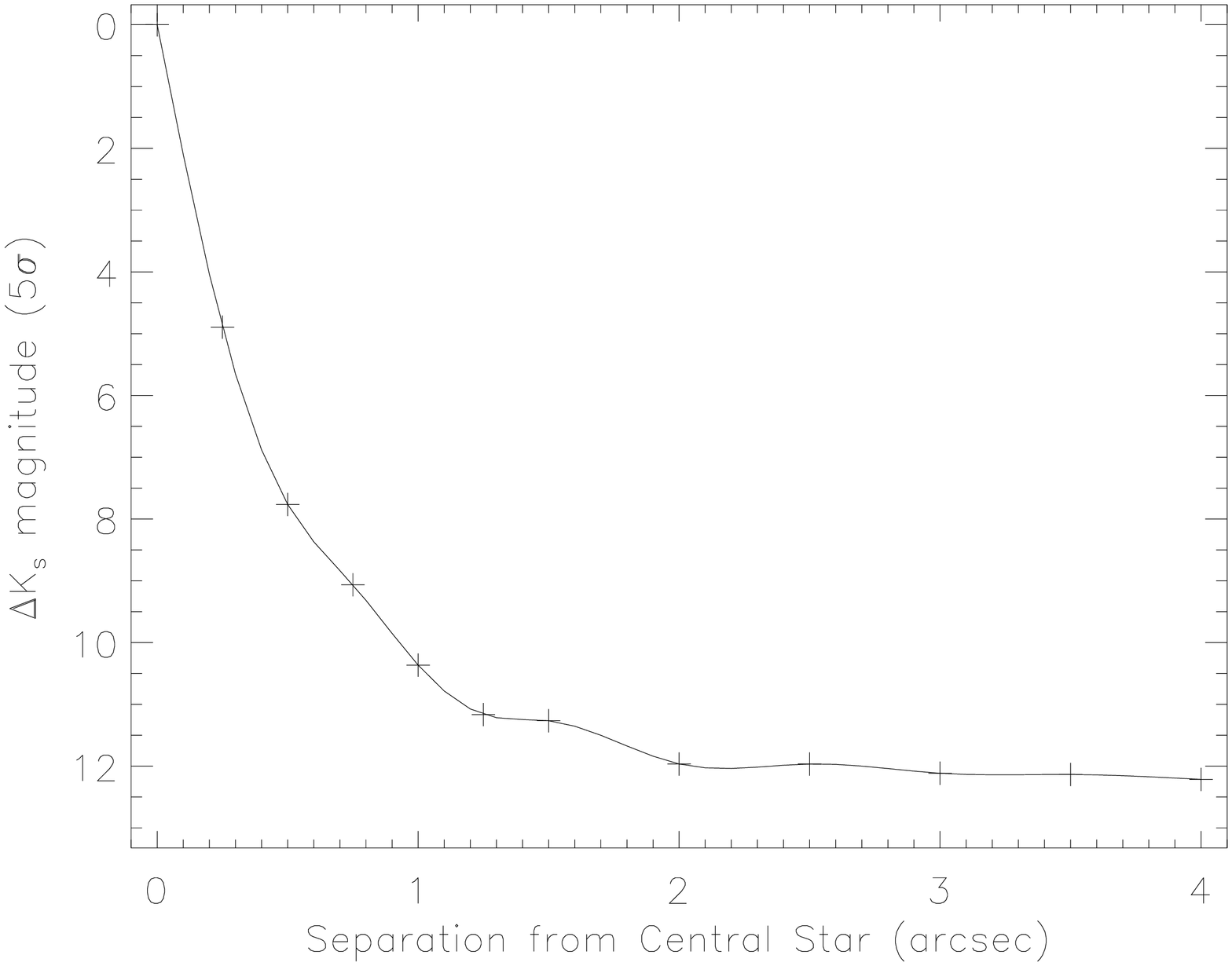}
\caption{Instrument sensitivity curve showing 5$\sigma$ $\Delta\mathrm{K_s}$ detection vs. distance from central star. The $15\!\times\!30\mathrm{\,s}$ deep $\mathrm{K_s}$ images of LP\,756-3 were used to insert unsaturated scaled-down images of the same star (example shown in Figure \ref{fig7}) and the S/N was measured at several locations in the image.
\label{fig8}} 
\end{figure}

An instrument sensitivity estimation using modeled faint companions was made. Unsaturated images of LP\,756-3 were scaled down and inserted into the fully reduced deep $\mathrm{K_s}$ image of the same star, which was considered to be a representative medium brightness star from our sample (2MASS $\mathrm{K_s}=8.435$, $d=15.7\mathrm{\,pc}$; Scholz et al. 2005). These ``fake planets'' were adjusted in brightness to simulate a 5$\sigma$ detection in the unsharp masked image (an example is shown in Figure \ref{fig7}). To determine the detection limit with sufficient accuracy, 5 fake planets were distributed randomly on a circle drawn around the primary. For each of these positions, 4 to 6 noise measurements from a $3\times3$ pixel region close to the companions' positions were taken and averaged. The final S/N was calculated by averaging the individual ratios of peak and noise count. Figure \ref{fig8} shows the sensitivity of our observation to the detection of faint close-in companions. Plotted are the maximum magnitude difference in $\mathrm{K_s}$ band vs. distance to the central star in order to detect faint companions at the 5$\sigma$ level.

\begin{figure}[tb]
\includegraphics[angle=90,width=\columnwidth]{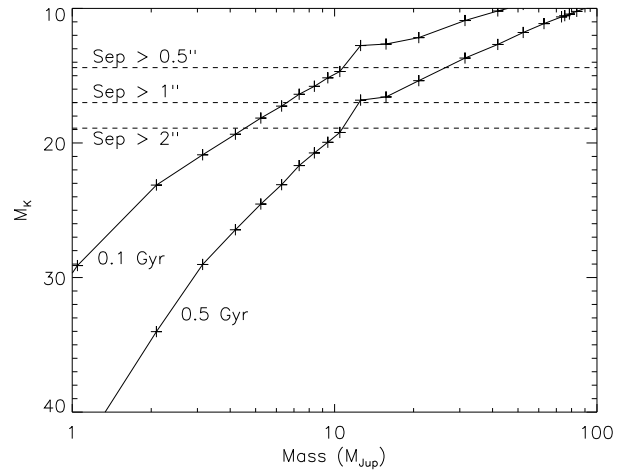}
\caption{COND isochrones according to the models of \citet{bar03} for absolute K magnitude vs. mass (in Jupiter-masses) for ages of 100 and 500\,Myr. The horizontal dashed lines show the detection limits of close in companions at the $5\sigma$ level derived from reference star LP\,756-3 at separations of $0\farcs5$, $1\arcsec$, and $2\arcsec$.
\label{fig9}} 
\end{figure}

Our instrumental detection limits predict that objects with $\Delta\mathrm{K_s}\approx7.8$ can be detected at separations $\gtrsim\!0\farcs5$ to the central star and objects with $\Delta\mathrm{K_s}\approx12$ at separations of $\gtrsim\!2\arcsec$. With LP\,756-3 as a reference, the absolute magnitudes of these hypothetical companions are M$_\mathrm{K_s}$ $\sim\!15.2$ and $\sim\!19.4$, respectively. All of our single stars are believed to have distances less than 20\,pc to the Sun. Accordingly, conservatively assuming an age of 500\,Myr, the models of \citet{bar03} translate the detection limit to: $\sim\!24\mathrm{\,M_{Jup}}$ at separations $\gtrsim\!10\mathrm{\,AU}$ and $\sim\!10\mathrm{\,M_{Jup}}$ at separations $\gtrsim\!40\mathrm{\,AU}$ (see Figure \ref{fig9}). Hence, we can exclude the existence of objects above these limits around the objects in Table \ref{tab1}.
We are using these evolutionary models as a convenient guide, although, there are likely uncertainties associated with these models at the lowest temperatures and hence our sensitivities (in terms of the lowest masses) should be viewed as estimates only.
Furthermore, the mass-luminosity relation of young planetary mass objects depends strongly on the exact age. The assumed age of 500\,Myr serves as a conservative limit estimate, since our objects are probably less than 300\,Myr old. The models of \citet{bar03}, calculated for ages 100 and 500\,Myr (Fig. \ref{fig9}), show a maximum difference of $\sim$ 9\,mag in the range of the lowest masses important for these observations. Hence, objects comparable to $300$\,Myr or even younger might be substantially brighter than the 500 Myr models predict and therefore the mass limit would be lower than our estimated values above. 

\section{Results}\label{ana}

\subsection{Notes on the Individual Multiples}
\subsubsection{G\,172-1}
In our search for planetary companions we come to the same result as \citet{zuc04b} of no faint companion candidates around G\,172-1. However, we were easily able to resolve this $0\farcs426$ binary for the first time into two almost equal-magnitude components.

\subsubsection{G\,274-24}
This binary was first observed by \citet{jao03} in 1999 as part of a triple system with a very wide B component with $37\farcs84\pm0\farcs05$ separation to the primary. This corresponds to a separation of $477\pm87 \mathrm{\,AU}$ when using our distance estimate of $12.6\pm2.3\mathrm{\,pc}$. We were not able to image the wide component due to our limited field of view of $30\arcsec\times30\arcsec$.

\subsubsection{G\,39-29}
G\,39-29 was discovered to be a binary by \citet{beu04}. Interestingly, there seems to be significant observable dynamics to the system since their measurement of $0\farcs54$ in September 2002 differs strongly from our $0\farcs783 \pm 0\farcs002$ in October 2005, which corresponds to a tangential motion in the plane of the sky of $\sim\!81\mathrm{\,mas/yr}$ relative to the primary. At the same time the position angle stayed approximately the same ($299^\circ$ in 2002, $300.6^\circ$ in 2005). We therefore conclude that the system is probably highly inclined. Further observations will be needed to set constraints on orbital elements.

\subsubsection{LP\,717-36}
We are the first to resolve this 0.53 arcsecond binary. A recent distance estimate by \citet{sch05} using low resolution spectroscopy resulted in a value of $10.4\mathrm{\,pc \pm 20\%}$ for the unresolved object. Our value of $20.2\pm4.7\mathrm{\,pc}$ agrees with that value on the $\sim\!1\sigma$-level, taking the newly discovered binarity into account by multiplying the \citet{sch05}-result by a factor of $1.4$.

\subsubsection{GJ\,2060}
The binarity of this highly interesting variable object with the alternative identifier V372\,Pup was discovered by \citet{zuc04b}. Unfortunately, separation and position angle are not published, which would have allowed further analysis of this close $2.8\pm0.5\mathrm{\,AU}$ projected separation (this paper) binary.

A trigonometric parallax was measured by the {\it Hipparcos} satellite \citep{per97}. Its value of $64.24\pm2.68\mathrm{\,mas}$ corresponding to a distance of $15.57\pm0.65\mathrm{\,pc}$ is in excellent agreement with our value of $15.8\pm2.3\mathrm{\,pc}$.

\subsubsection{2MASS10364483+1521394}
This object (subsequently 2M1036) is found to be a triple system. It shows a $\mathrm{H}=8.7$, K$\mathrm{_s}=8.5$ primary component accompanied by a pair of equal magnitude ($\mathrm{H}=9.9$, K$\mathrm{_s}=9.6$) objects at $1\farcs06$, which themselves are separated by $0\farcs189$ (see Figure \ref{fig5}).

\subsubsection{LHS\,2739}
With a distance of $30.2\pm7.7\mathrm{\,pc}$, this high proper motion system was calculated to have the second-largest distance of all objects in our sample. Its binary nature is reported here for the first time.

\subsubsection{2MASS13345147+3746195}
This binary (hereafter 2M1334) was discovered with a separation of $0\farcs082$, just at the resolution limit of this survey of $\sim\,0\farcs08$.

\subsubsection{G\,180-11}
After its binarity was first published by \citet{mcc01}, the primary was discussed by \citet{cru03} as hosting either an 11,000\,K white dwarf or an M6 companion. From our observed $\Delta\mathrm{K_s}=1.97$ and $\mathrm{H}-\mathrm{K_s}=0.24$ for the companion we can exclude the possibility of a white dwarf companion, since we would expect $\mathrm{H}-\mathrm{K_s}$ to be less than zero for a predicted 11,000\,K white dwarf according to models of \citet{ber95}. We therefore conclude the companion to be a late type main sequence star. Our spectral type estimations yield a value of M$7\pm2$ for the secondary, which is therefore the latest type companion in this sample.

The inferred mass of the secondary is possibly below the hydrogen burning limit ($M_\mathrm{B}=0.076^{+0.058}_{-0.025}\,M_\sun$), which makes this object a brown dwarf candidate. This is also the lowest mass companion found in this survey.

\subsubsection{LP\,331-57}
\citet{sch05} found a photometric distance of $12.7\mathrm{\,pc}\pm20\,\%$ for the unresolved binary. This is in good agreement with our value of $18.0\pm1.2\mathrm{\,pc}$ when multiplying with a factor of $\sqrt{2}$ to account for the binarity discovered by our survey. 

\subsubsection{Steph\,2018}
The binary nature of this $1\farcs571\pm0\farcs003$ separated slow-moving system was first found by \citet{mcc01}.

\subsubsection{LP\,642-48}
We were able to clearly resolve the components of this newly discovered tight $0.099\arcsec$ binary. 

\subsubsection{LP\,764-40}
Our observations are the first to identify this system as a binary, with separation 1.9 arcseconds. The optical spectrum \citep{bee96} shows strong Balmer emission, consistent with the young age inferred from the strong X-ray flux.

\subsection{Rejected Very Faint Companion Candidates}
Our observations reveal a number of faint sources near the targeted stars that we have rejected as potential companions. Here, we outline the rational behind those decisions for a selection of objects.

\subsubsection{Candidate Companion to G\,173-39 is a Background Object}\label{G173-39}
\begin{figure}[tb]
\includegraphics[angle=0,width=\columnwidth]{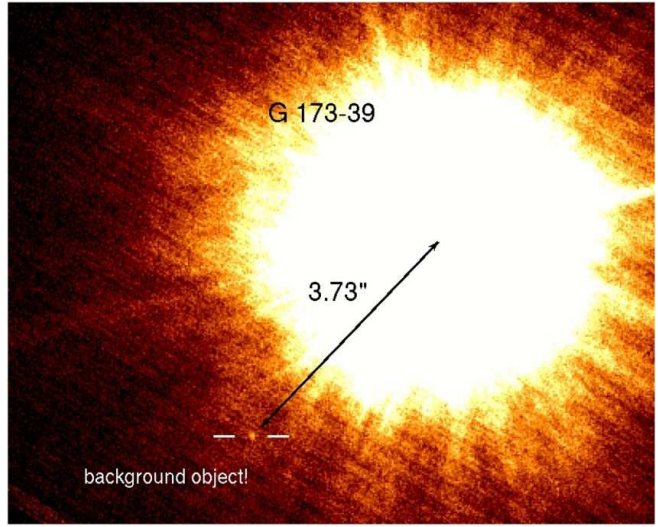}
\caption{Faint $\Delta \mathrm{K_s}=12.16$ and $\Delta \mathrm{H}=12.4$ companion to G\,173-39 was detected with a separation of $3\farcs73$ to the central star. Assuming that this object resides in the same distance as G\,173-39 (14.3\,AU; Gliese \& Jahrei{\ss} 1991), this corresponds to a $\sim 10\mathrm{\,M_{Jup}}$ companion in 53.3\,AU separation to the central object at an age of 500\,Myr. Follow-up observations revealed its background nature through proper motion arguments.
\label{fig10}} 
\end{figure}

A very faint $\Delta\mathrm{H}=12.4$ and $\Delta\mathrm{K_s}=12.2$ point-like object was found at a position angle of $134.5^\circ$ at a distance of $3.73\arcsec$ to the primary (see Figure \ref{fig10}). Follow-up observations with the Gemini North Telescope were conducted 2 months after first detection to test for proper motion correlation of the primary and companion candidate. From the two epoch data a relative motion of the central object and companion was derived and compared with the known proper motion of $0.37\mathrm{\,arcsec/yr}$ of the primary. We find that the data are consistent with a distant background object.

\subsubsection{Candidate Companion to LP\,390-16 is a Background Object}
We detected a companion to LP\,390-16 at only $2\farcs81$ separation with a delta magnitude of more than 7\,mag in H and K band. Combining proper motion with point source locations in old exposures from POSS\,I, however, we find the object to be consistent with a non-moving object, and therefore conclude it to be a background star.

\subsubsection{Candidate Companions to G\,125-15 are Background Objects}
Follow-up observations helped to reject 7 companion candidates at separations between $2\farcs9$ and $6\farcs7$ to G\,125-15. After a first observation in June 2005 the central object moved $0.16\arcsec$ due to proper motion until a second observation in April 2006. This corresponds to $\sim 5.7\mathrm{\,pixels}$ on the detector, which exposes the fixed companion candidates to be distant background stars.

\subsubsection{Candidate Companions to LP\,756-3 are Background Object}
In $5\farcs8$ separation to the central object we detected two faint objects with a separation of $0\farcs28$. The proper motion of LP\,756-3 (0.353 arcsec/yr) and second epoch data 10 months after our first observation were used to reveal their background nature.

\subsection{Are the Companions Physically Bound to the Primaries?}
Four of the 13 observed multiples already have published separations and position angles; proper motions are known for all. Comparing our data taken in 2005 with the earlier epoch data of the other groups we find that all four binaries are consistent with physically related companions.

Evidence for the physical nature of the other multiples is given by proper motion arguments. Our sample consists of relatively nearby M stars that show significant proper motions. Looking at multiple epoch data, a background star mimicking a bound companion would remain fixed while the foreground object would move by. As the $\Delta\mathrm{mag}$ of all observed binaries range from $0$ to $\sim\!2$ only (implying apparent magnitudes of $6.5\lesssim\mathrm{K_s}\lesssim10$), these bright companions would show up in POSS\,I data taken at much earlier epoch. No bright background objects were found in the vicinity of any of the estimated secondary locations leading to the conclusion that these cool companions are physically bound to their primaries.

\section{Discussion}\label{dis}
Observations of binaries and their inherent parameters are important for setting constraints on the formation mechanism of stars in general. This sample of young, early-M dwarfs allows insight into a population of stars in between the well-known G and K star population and the distinctly different VLM population \citep{bur06}, which became analyzable only a few years ago with the development and employment of highly sensitive AO and the {\it HST} (Close et al. 2003; Gizis et al. 2003; Bouy et al. 2003).

\subsection{The Multiple Star Fraction}
We found 12 binaries and 1 triple system in our sample of 41 objects. In order to derive a multiple star fraction free of Malmquist bias, we consider only the systems in our sample whose distances are within 20\,pc. Eight of the observed multiples are likely within 20\,pc of the Sun, as are--estimatedly--all of the observed single stars. However, the distances of two of the 8 objects (G\,172-1, 2M1036) are within $1\sigma$ of the 20\,pc limit, leaving a rather large probability that they are not part of the defined volume. Likewise, two binaries have estimated distances $<1\sigma$ greater than 20\,pc (LP\,717-36 and LP\,642-48), so they may be part of the limited sample. This therefore leaves $8^{+2}_{-2}$ confirmed binaries within 20\,pc. To eliminate uncertainties caused by the distance dependence of the sensitivity, we conservatively cut the sample at a mass ratio of $q=0.1$. There are 28 objects in this sub-sample other than the confirmed binaries. Four of these have candidate companions with $q<0.1$. Hence, we include them in this binary fraction as single stars. In addition, we include one possible binary system in the upper limit estimate. This results in a total of $8^{+3}_{-2}$ multiples in a sample of $35^{+2}_{-2}$ objects (upper and lower limits correspond to each other, respectively), resulting in a nominal binary fraction of $23^{+15}_{-9}\,\%$ for $q>0.1$ and $\mathrm{sep.}>1.6$\,AU (we use binomial statistics to estimate the uncertainty as described in \citet{bur03}). The cut in separation is determined by the minimum resolution of $\mathrm{FWHM}=0\farcs08$ at the distance limit of 20\,pc. We note, however, that this above examination of the binary fraction does not account for selection effects in the sample selection.

\subsection{The Mass Ratio Distribution}\label{massratio}
\begin{figure}[tb]
\includegraphics[angle=90,width=\columnwidth]{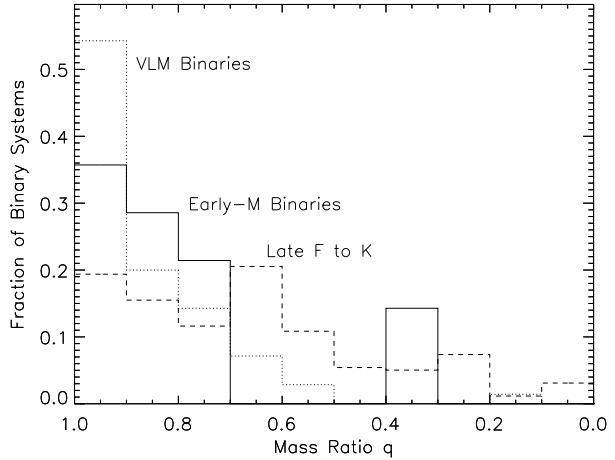}
\caption{Mass ratio $q=M_B/M_A$ distribution. The continuous line shows the distribution derived from the 14 early-M binaries in this paper (triple 2M1036 included as two binaries, see \S \ref{massratio}). The dashed line shows the distribution of binaries included in an early F to K star sample by \citet{rei02} that have separations $<\!100\mathrm{\,AU}$. The dotted line shows the $q$ distribution for VLM stars from \citet{bur06}. While the F to K binary distribution is described by a rather flat distribution, slightly more populated at higher $q$, the VLM distribution shows a significant peak for equal mass binaries. The early-M binary distribution with its broad peak at $1 \ge q > 0.7 $ might represent an intermediate distribution between the earlier and later types. Due to the high sensitivity of our measurements, only the last, $q<0.1$ bin is significantly affected by incompleteness.
\label{fig11}} 
\end{figure}

Figure \ref{fig11} shows the mass ratio $q=M_B/M_A$ distribution derived from the 13 observed multiples using the masses listed in Table \ref{tab3}. The triple system 2M1036 is accounted for as two binaries, the binary BC and the A(BC) system. Mass ratios for these two objects are therefore $q_\mathrm{BC}=M_\mathrm{C}/M_\mathrm{B}$ and $q_\mathrm{A(BC)}=(M_\mathrm{B}+M_\mathrm{C})/M_\mathrm{A}$. In the following paragraphs we sometimes refer to 14 ``binaries'', but one should keep in mind that this means 12 binaries and 1 triple system.

\begin{figure}[tb]
\includegraphics[angle=90,width=\columnwidth]{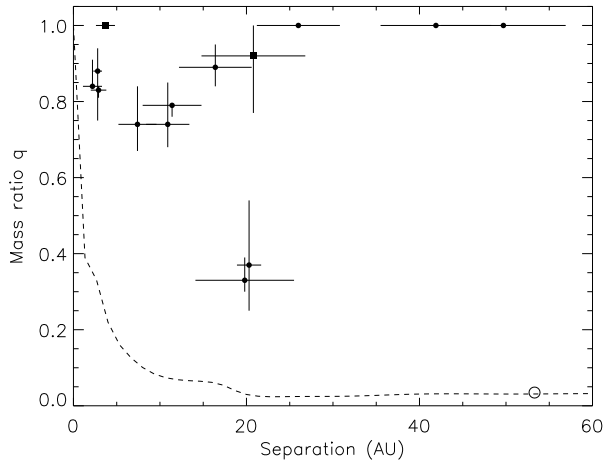}
\caption{Mass ratio $q=M_B/M_A$ as a function of projected binary separation in astronomical units for the 12 discovered binaries (filled circles). The triple 2M1036 is included as the two binaries A(BC) and BC (squares; for further explanation see \S \ref{massratio}). The dashed line represents the sensitivity limit of our survey, derived from the sensitivity relation shown in Figure \ref{fig3} and 500\,Myr models from \citet{bar03}. As a demonstration of detection capabilities, the open circle at $\mathrm{sep.}=53.3\mathrm{\,AU}$ and $q=0.035$ is the companion seen in Figure \ref{fig10} if it were a physical companion to G\,173-39.
\label{fig12}} 
\end{figure}

Twelve out of the 14 have mass ratios of $q\gtrsim0.74$; we regard this as evidence that equal-mass systems are preferred for young, early-M dwarf primaries. Our observations are sensitive to binaries with mass ratios $q<0.1$ for separations greater than $\sim\!8\mathrm{\,AU}$ (see Figure \ref{fig12}). Observations have shown that there is a significant lack of very low-mass objects around main sequence stars for separations smaller than 5 AU (the ``brown dwarf desert''; Marcy \& Butler 2000). We therefore do not expect to have missed a large quantity of objects. However, we have to assume that for $q<0.1$ the sample is likely incomplete.

This emphasis on $q\sim1$ is not seen in the distributions of earlier type stars. \citet{rei02} show that for late F to K binaries with separations of less than 100\,AU the distribution of mass ratios is rather flat with only marginal tendency towards equal mass binaries. For field VLM binaries, however, the opposite effect is observed; recent statistical evaluations of all currently known VLM binaries \citep{bur06} show a significant peak at $q\!\sim\!1$. It is therefore possible that the mass ratio distribution of our early-M binaries sample has an intermediate distribution between the flat F-K distribution and the strongly peaked field VLM distribution. 

However, our sample of binaries is rather small. Further observations of a larger number of young M stars will be needed to show if this relation holds.

\subsection{Mass Ratio vs. Separation}
Mass ratios $q=M_\mathrm{B}/M_\mathrm{A}$ as a function of binary separation are shown in Figure \ref{fig12}. While at separations $\lesssim 20$\,AU systems are distributed over a broad range of mass ratios, all three binaries with separations $> 25$\,AU show mass ratios of $q=1$. The sensitivity of the observations would have allowed for the detection of much lower mass ratios at these separations (see Figure \ref{fig10} for example), indicating a real lack of objects with low mass ratios and high separations. However, a sample of 14 binaries cannot fill the accessible space in the $(q,\mathrm{separation})$ plane uniformly, limiting the significance of this result due to low number statistics.

\subsection{The Separation Distribution}
\begin{figure}[tb]
\includegraphics[angle=90,width=\columnwidth]{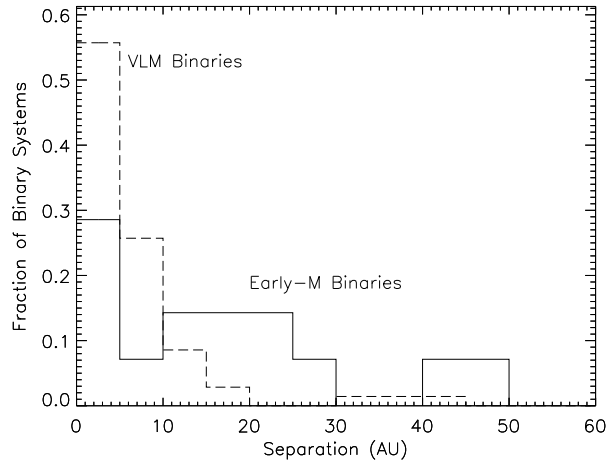}
\caption{Distribution of projected binary separations. The solid line shows the separation distribution based on our early-M sample consisting of 12 binaries and a triple system counted as two binaries A(BC) and BC. The dashed line describes the projected separation/semi major axis distribution of 70 VLM binaries \citep{bur06}. The $\mathrm{sep.}<5\mathrm{\,AU}$ bin is probably incomplete due to the finite resolution of $0\farcs08$ corresponding to 1.6\,AU in 20\,pc.
\label{fig13}} 
\end{figure}

Figure \ref{fig13} shows the separation distribution of our set of 14 binaries. The physical projected separations range from 2.2 to 49.7\,AU and possess a distribution that shows no significant peak. However, since the observations cover separations up to 15 arcseconds, which corresponds to 300\,AU at a distance of 20\,pc, we conclude that the decreasing number of detected objects for separations higher than $\sim\!50$\,AU is a real effect. The underlying distribution therefore probably peaks within 0 to 50\,AU. A least-$\chi^2$ fit of a Poisson distribution to the data suggests a peak at $13^{+14}_{-9}\mathrm{\,AU}$. This is consistent with earlier results, for instance by \citet{fis92}, who report a broad peak at $4$--$30\mathrm{\,AU}$ for M2--M4.5 binaries.

The separation distribution, again, shows evidence that young, early-M binaries possess transitional characteristics between earlier and later type populations. With a distribution peaking at $\sim\!30\mathrm{\,AU}$ \citep{duq91}, G stars seem to have---on average---greater separations than early-M stars. Field VLM binaries, however, are preferably found in significantly tighter systems. The distribution of 70 binaries with spectral types later than M5.5 shows a significant peak at $3$--$10\mathrm{\,AU}$ \citep{bur06} with a steep quantitative decrease of systems with higher separations.

G--K binaries are described by a mass ratio distribution and a separation distribution that are very different from the VLM binaries' characteristics; young, early-M binaries appear to show an intermediate behavior. The higher and the lower mass distributions are apparently connected through a smooth transition embodied by early-M binaries.

\section{Summary}
We have conducted a survey of 41 nearby, early-M stars of ages comparable to the Pleiades cluster, searching for companions down to planetary masses using the Gemini North telescope with its AO system Altair. Thirteen stars were observed to host fairly bright companions, 8 of these multiples (7 binaries, 1 triple) are new discoveries. We derived physical parameters for each of the multiple components allowing for further analysis of the early-M star population confirming existing statistical results and contributing additional empirical constraints on star and binary formation.

The binary fraction of this sample of young, early-M stars was found to be $23^{+15}_{-9}\,\%$ for mass ratios $q>0.1$ and $\mathrm{sep.}>1.6$\,AU. However, a full analysis of systematic errors and biases could not be made, this value might therefore differ significantly from the actual value for this group of stars.

The mass ratio distribution was analyzed and compared with distributions drawn from F--K stars and from VLM stars. The young, early-M binary data was shown to be best described as an intermediate distribution between the almost flat early type star distribution and the strongly $q\rightarrow 1$ peaked field VLM distribution, however, our sample size is relatively small.

The most likely separation was found to be $13^{+14}_{-9}\mathrm{\,AU}$, which is less than the median value of the G binary distribution. Field VLM binaries, however, tend to be bound in significantly tighter systems. We conclude that early-M binaries possess characteristics filling the gap between the very distinct field VLM population and stars earlier than M.

A sensitivity estimate to the detection of planetary mass companions was made using planetary evolution models. Since we found no faint physical companions around 37 of the observed sources we can exclude---conservatively assuming an age of 500\,Myr---companions more massive than 10\,M$_\mathrm{Jup}$ at separations of $\gtrsim\!40\mathrm{\,AU}$ and heavier than $23\mathrm{\,M_{Jup}}$ at separations of $\gtrsim\!10\mathrm{\,AU}$ around all these objects.

\acknowledgements
We thank the anonymous referee for many valuable comments leading to an improved paper.
The authors are grateful to Jean-Ren{\'e} Roy, Deputy Director \& Head of Science at Gemini North, for kindly allocating special DDT time for follow-up observations.
This work is based on observations obtained at the Gemini Observatory, which is operated by the Association of Universities for Research in Astronomy, Inc., under a cooperative agreement with the NSF on behalf of the Gemini partnership: the National Science Foundation (United States), the Particle Physics and Astronomy Research Council (United Kingdom), the National Research Council (Canada), CONICYT (Chile), the Australian Research Council (Australia), CNPq (Brazil) and CONICET (Argentina).
Program IDs of the Gemini observations are GN-2005A-Q-28, GN-2005B-Q-14, and GN-2006A-Q-83.
This publication makes use of data products from the Two Micron All Sky Survey, which is a joint project of the University of Massachusetts and the Infrared Processing and Analysis Center/California Institute of Technology, funded by the National Aeronautics and Space Administration and the National Science Foundation.
This research has made use of the SIMBAD database, operated at CDS, Strasbourg, France.
The National Geographic Society - Palomar Observatory Sky Atlas (POSS-I) was made by the California Institute of Technology with grants from the National Geographic Society.
LMC is supported by an NSF CAREER award and the NASA Origins of Solar Systems program.

\end{document}